\documentclass[letterpaper,12pt]{article}   
\usepackage{osajnl2} 
\usepackage[draft]{hyperref} 

\begin{document}

\title{Truly random number generation via entropy amplification}


\author{Yu Liu, Mingyi Zhu, and Hong Guo$^{*}$}
\address{
CREAM Group, State Key Laboratory of Advanced Optical Communication \\ Systems and Networks (Peking University) and Institute of Quantum Electronics,\\ School of Electronics Engineering and Computer Science, Peking University, Beijing 100871, China\\
$^*$Corresponding author: hongguo@pku.edu.cn
}

\begin{abstract}We present a simple setup to implement truly random number generator based on the measurement of the laser phase noise.
From the entropy point of view, we estimate the number of truly random bits that can be extracted from the sampled Byte. With a simple method of adopting the
$m$-least-significant-bit, we amplify the entropy of the original bit sequence and realize a truly random bit generation rate of $300$ Mbps.
\end{abstract}


\maketitle 

\noindent The application of random number generator (RNG) covers wide areas of cryptography, statistical sampling, computer simulations, etc.
Traditionally, RNG is implemented using the algorithm-based pseudo-random number generator (PRNG), while in most application areas, PRNGs are not enough.
So far, it is commonly accepted that pure quantum process is the only solution for truly random number generator (TRNG).
Recently, two kinds of quantum noise of laser, phase noise and relaxation oscillation noise, are chosen to generate truly random bit sequence
with high generation rate \cite{TWZ,QI,U08,R09}.
Phase noise, directly measured using self-delayed homodyne detection \cite{TWZ,QI}, can be used to generate random bits with good
randomness, but the final random number generation rate is limited by the laser linewidth;
relaxation oscillation noise, amplified by chaotic behavior, can be used to achieve ultra-high random bit generation rate \cite{U08,R09},
but the random bits generated by chaotic laser involve more classical noise (not truly random)
and so cannot have good results in three-sigma randomness tests \cite{TWZ,R09}.
To generate an original random bit sequence, either an oscilloscope \cite{QI} or an analog-digital-converter (ADC) \cite{R09} is available
for high speed multi-bit per measurement data recording, but due to data size limitation in one scan process, using oscilloscope will reduce the generation rate for long random bit sequence.
Multi-bit per measurement method increases the random number generation rate but introduces additional correlations. The proper number of bits extracted from the recorded data can be estimated by randomness evaluation.

In this Letter, we propose a simple method to evaluate the randomness from the entropy point of view, which gives the upper and lower estimations of the maximal bits that can be extracted from one measurement.
The laser phase noise (Gaussian random variable \cite{L67}) measurement based TRNG scheme is adopted,
where a single-mode vertical cavity surface emitting laser (VCSEL) is used and sampled by an ADC. After entropy amplification by bit extraction,
we realize a $300$ Mbps TRNG, which passes three standard statistical tests. Note that this bit extraction method is also applicable for other random sources.

\begin{figure}[t]
\centerline{
\includegraphics[width=8.3cm]{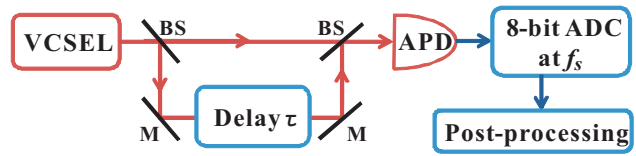}}
 \caption{(Color online) Schematic setup of TRNG based on the laser phase noise measurement using delayed self-homodyne
method. BS: beam splitter; APD: Silicon avalanche photodetector with the low (high) cutoff frequency of 50 kHz (1 GHz); ADC:
8-bit binary analog-digital-converter working at $f_s=100$ MHz.}\label{setup}
\end{figure}

The schematic setup is shown in Fig. \ref{setup} (see also \cite{TWZ}) and the delayed self-homodyne method is used to measure the
phase noise of the VCSEL. The beat signal of $E(t)$ and $E(t+\tau)$ is detected by the avalanche photodetector (APD) and sampled
at $f_s=100$ MHz by the 8-bit binary ADC.
When the delay time is much longer than the coherence time of laser ($\tau\gg\tau_{\rm coh}$),
the autocorrelation function of the laser electric field is eliminated \cite{H82}, i.e.,
$\langle E^*(t)E(t+\tau)\rangle \propto \exp(-|\tau|/\tau_{\rm coh})\rightarrow 0$,
where $\tau_{\rm coh} = (\pi\Delta\nu_{\rm Laser})^{-1}$ and $\Delta\nu_{\rm Laser}$ is the laser
linewidth. In this case, the beat signal of these two mutually independent laser fields can be considered as a good candidate to implement TRNG.

\begin{figure}[t]
\centerline{
\includegraphics[width=6.5cm]{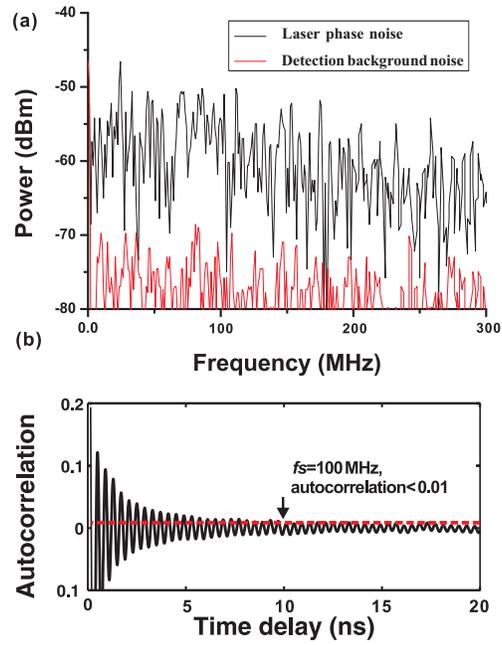}}
 \caption{(Color online) (a) The laser phase (detection background) noise of the laser field is observed at $1.4$ mA ($0$ mA), while the threshold
current is $1.3$ mA. (b) Autocorrelation function of the beat signal vs time delay (red line corresponds to autocorrelation of $0.01$). The sampling frequency is $f_s = 100$ MHz ($t_s = 10$ ns).}
\label{spectrum}\end{figure}

Since the phase noise of the laser is inversely proportional to the laser power, while the classical noises are laser power independent \cite{Vahala},
we choose working current just above the threshold current to ensure that the phase noise is dominant and the true randomness can thus be guaranteed.
In experiment, a $795$ nm VCSEL laser works at 1.4 mA (threshold current 1.3 mA),
and its linewidth is $\Delta\nu_{\rm Laser} = 120$ MHz ($\tau_{\rm coh} = 2.65$ ns).  With delay
time $\tau=5$ ns, the power spectral density of the beat signal is shown in Fig. \ref{spectrum} (a).
In Fig. \ref{spectrum} (b), we calculate the autocorrelation function $R_{beat}(t)$ from the power density spectrum in Fig. \ref{spectrum} (a) using Wiener-Khintchine theorem \cite{W30}.
Without any post-processing, the autocorrelation of the laser source sampled by 8-bit ADC at $100$ MHz (time delay $10$ ns) is smaller than $10^{-2}$.

\begin{figure}[t]
\centerline{
\includegraphics[height=9cm]{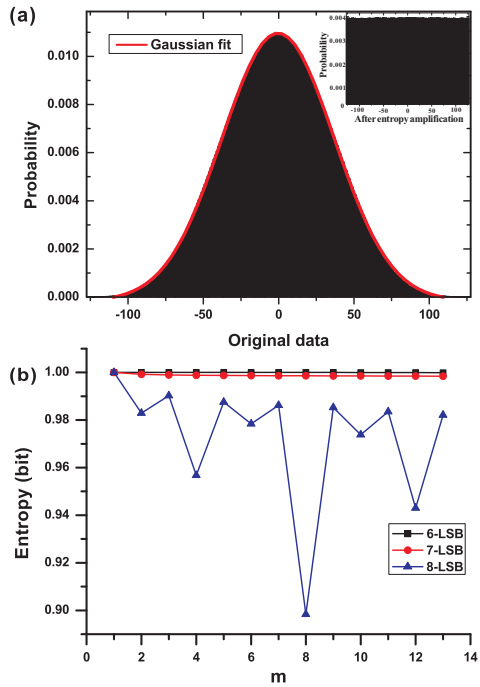}}
 \caption{(Color online) (a) Histogram of the laser intensity digitized by an 8-bit ADC after reducing bias by exclusive-OR post-processing, and
after entropy amplification by bit extraction (inset). (b) Normalized entropy $H(m)/m$ of the random bit sequence when the $m$-least-significant-bit ($m$-LSB) is adopted.}\label{histogram}
\end{figure}

Due to the dependence on the external parameters in the experiment, the sampled original data digitized by an 8-bit ADC is biased. To reduce the bias, we perform a simple exclusive-OR post-processing between every two 8-bit original data, which is a common method to improve the randomness of RNG \cite{XOR}.
After the bit extraction, the histogram of the data is shown in Fig. \ref{histogram} (a), well fitted with a Gaussian profile.
Since truly randomness should have a uniform distributed histogram with Shannon entropy of 8 bits per Byte, the randomness in this original data is limited by the correlation within the Byte. To evaluate the randomness, we calculate Shannon entropy using the experiment results (histogram data): $H_{S}=-\sum_{i=-128}^{127} p_i\log_2{p_i}=7.1925$, where $p_i$ is the probability of the data in the $i^{th}$ digitized bin (totally $2^8$ bins). This can roughly be viewed as that this data can extract $7.1925$ random bits out of every Byte using appropriate, yet maybe complicated bit extraction post-processing. To assure that Shannon entropy can evaluate the randomness of the source, we calculate the entropy of laser source according to its Gaussian fitting curve (Fig. \ref{histogram} (a)) using $H_S'=\frac{1}{2}\log_2(2\pi\rm{e}\sigma^2)$, where $\sigma$ is the standard deviation of Gaussian profile. The result of the entropy of the laser source is $H_S'=7.2685$ bits per Byte, and experiments show these two values of Shannon entropy get closer when the size of the data is larger.

To avoid complicated post-processing, adopting $m$-least-significant-bit ($m$-LSB) out of each Byte is an efficient method to amplify the randomness. But this method is not strong in eliminating the correlation, and extracting $7$-LSB according to Shannon entropy cannot pass the random tests. Therefore, we change the randomness indicator to min-Entropy, $H_{min}=-\log_2\{\max(p_i)\}$, which uses only the maximal probability and represents the worst-case scenario \cite{minEn}. By simply adopting the last 6-bit in each Byte according to the result, $H_{min}=6.5083$ bits per Byte, the histogram of the new random bit sequence shows good uniformity (inset of Fig. \ref{histogram} (a)) and represents good randomness. Hence, we can consider Shannon entropy and min-entropy, respectively, as the upper and lower estimations of the randomness of the laser source.

For true randomness, Shannon entropy and min-entropy have the same value and both equal $8$ bits per Byte. In our experiments, after the bit extraction process, the values of these two entropies are amplified to $7.99996$ and $7.98333$ bits per Byte, respectively. To further illustrate the entropy amplification, we calculate the normalized entropy $H_{\rm nor}=H(m)/m$ of the random bit sequence before and after bit extraction (Fig. \ref{histogram} (b)), where $H(m)$ represents Shannon entropy with the length of tested-bit sequences $m$. Since Shannon entropy holds $H(m)=m$ for true randomness, normalized entropy should be unity for any $m$. It can be seen from Fig. \ref{histogram} (b) that, after bit extraction, the entropy of the random bit sequence is amplified and gets much closer to unity.

\begin{figure}[t]
\centerline{
\includegraphics[width=6.3cm]{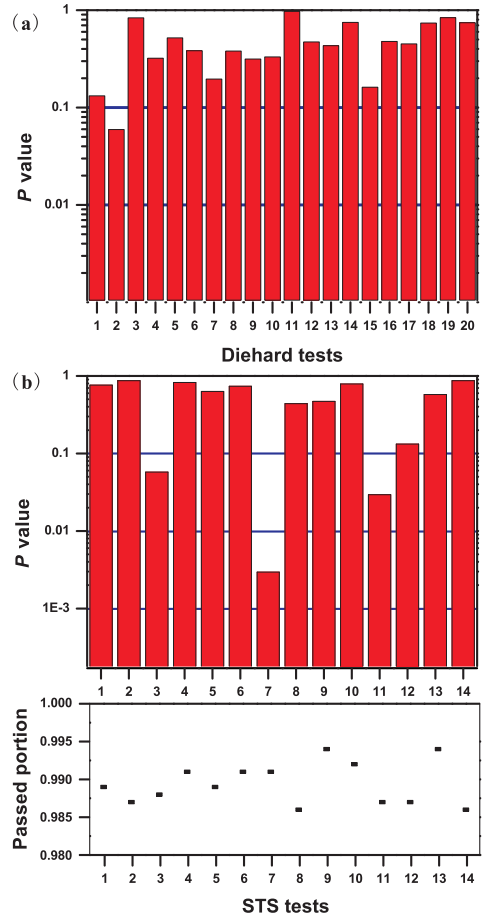}}
 \caption{(Color online) The results of the randomness tests of (a) Diehard and (b) STS. The Diehard test is considered successful when all $P$ values satisfy $0.01<P<0.99$. The STS test is considered successful when all $P$ values are larger than 0.0001 and the passed portions are between $0.99\pm0.095$.}
\label{tests}\end{figure}

For randomness tests, we continuously record a final random bit sequence of 1 Gbit, and it passes all three standard random tests, i.e., ENT: a pseudorandom number sequence test program \cite{ENT}, Diehard: a battery of tests of randomness \cite{Diehard}, and NIST-STS: the National Institute of Standards and Technology-Statistical Test Suite \cite{STS}. The ENT results are: entropy $= 1.000000$ bit per bit (the optimum compression would reduce the bit file by $0\%$). $\chi^2$ distribution is $0.13$ (randomly would exceed this value less than $96.47\%$ of the times). Arithmetic mean value of data bits is $0.5000$ ($0.5 =$ random).
Monte Carlo value for $\pi$ is $3.141676275$. Serial correlation coefficient is $0.000041$. The results of Diehard and STS test are shown in Fig. \ref{tests} (a) and (b), respectively.

In summary, we implement a simple and compact TRNG based on the measurement of the laser phase noise, which passes all three standard statistical random tests.
To amplify the entropy, we estimate the maximal bits that can be extracted from each sampled Byte by LSB-adoption, and realize a $300$ Mbps truly random number generation rate.

This work is supported by the Key Project of National Natural Science Foundation of China (NSFC) (grant
60837004) and National Hi-Tech Research and Development (863) Program. We acknowledge the help from W. Tang, W. Wei, Y. Li, J. Zhou and J. Zhang.

\clearpage

\section*{List of Figure Captions}

Fig. 1. (Color online) Schematic setup of TRNG based on the laser phase noise measurement using delayed self-homodyne
method. BS: beam splitter; APD: Silicon avalanche photodetector with the low (high) cutoff frequency of 50 kHz (1 GHz); ADC:
8-bit binary analog-digital-converter working at $f_s=100$ MHz.
\\
\\
Fig. 2. (Color online) (a) The laser phase (detection background) noise of the laser field is observed at $1.4$ mA ($0$ mA), while the threshold
current is $1.3$ mA. (b) Autocorrelation function of the beat signal vs time delay (red line corresponds to autocorrelation of $0.01$). The sampling frequency is $f_s = 100$ MHz ($t_s = 10$ ns).
\\
\\
Fig. 3. (Color online) (a) Histogram of the laser intensity digitized by an 8-bit ADC after reducing bias by exclusive-OR post-processing, and
after entropy amplification by bit extraction (inset). (b) Normalized entropy $H(m)/m$ of the random bit sequence when the $m$-least-significant-bit ($m$-LSB) is adopted.
\\
\\
Fig. 4. (Color online) The results of the randomness tests of (a) Diehard and (b) STS. The Diehard test is considered successful when all $P$ values satisfy $0.01<P<0.99$. The STS test is considered successful when all $P$ values are larger than 0.0001 and the passed portions are between $0.99\pm0.095$.


\clearpage


\clearpage

\clearpage

\clearpage

\end{document}